\documentclass[aps,prd,twocolumn,superscriptaddress,nofootinbib,eqsecnum]{revtex4-1}

\pdfoutput=1

\usepackage{amsfonts}
\usepackage{amsmath}
\usepackage{amssymb}
\usepackage{graphicx,color}
\usepackage{float}
\usepackage{hyperref}
\usepackage{subfigure}
\usepackage{dcolumn}

\begin{document}

\title{Warm brane inflation with an exponential potential:
  A consistent realization away from the swampland}

\author{Vahid  Kamali}
\affiliation{Department of Physics, McGill University, Montreal, Quebec, H3A 2T8,
Canada}
\affiliation{Department of Physics, Bu-Ali Sina (Avicenna) University, Hamedan 65178, 016016, Iran}
\affiliation{School of Physics,
Institute for Research in Fundamental Sciences (IPM),
19538-33511, Tehran, Iran}

\author{Meysam Motaharfar}
\affiliation{Department of Physics, Shahid Beheshti University, G. C., Evin,Tehran 19839, Iran}

\author{Rudnei O. Ramos}
\affiliation{Departamento de F\'{\i}sica Te\'orica, Universidade do Estado do Rio de Janeiro, 
20550-013 Rio de Janeiro, RJ, Brazil }

\begin{abstract}

It has very recently been realized that coupling branes to higher
dimensional quantum gravity theories and considering the consistency
of what lives on the branes, one is able to understand whether such
theories can belong either to the swampland or to the landscape. In
this regard, in the present work, we study a warm inflation model
embedded in the Randall-Sundrum braneworld scenario. It is explicitly
shown that this model belongs to the landscape by supporting a strong
dissipative regime with an inflaton steep exponential potential. The
presence of extra dimension effects from the braneworld allow achieving 
this strong dissipative regime, which is shown to be both
theoretically and observationally consistent. In fact, such strong
dissipation effects, which decrease towards the end of inflation,
together with the extra dimension effect, allow the present
realization to simultaneously satisfy all previous restrictions
imposed on such a model and to evade the recently proposed swampland
conjectures. The present implementation of this model, in terms of an
exponential potential for the scalar field, makes it also a possible
candidate for describing the late-time Universe in the context of a
dissipative quintessential inflation model and we discuss this
possibility in the Conclusions.

\end{abstract}

\maketitle

\section{Introduction} 

{}Flattening, isotropizing, and homogenizing the Universe and
reproducing the adiabatic, nearly Gaussian, and quasiscale invariant
spectrum of primordial density fluctuations in accordance with the
observational cosmological data~\cite{Aghanim:2018eyx} require
implementing a mechanism complementing the standard big bang cosmology
to portray the very early Universe. There are today several scenarios
attempting to achieve all of these features, namely, ekpyrotic/cyclic
theories~\cite{Khoury:2001wf}, matter-bounce
scenario~\cite{Finelli:2001sr}, pre-big bang
cosmology~\cite{Gasperini:1992em}, string gas
cosmology~\cite{Nayeri:2005ck}, pseudoconformal
cosmology \cite{Hinterbichler:2011qk}, to name but a few. Yet,
inflation~\cite{inflation}, an exponentially accelerated expansion
driven by a scalar field $\phi$ rolling down a sufficiently flat region of a
potential $V(\phi)$, is still the simplest and most successful one
among all other alternatives. Despite its tremendous success,
inflation suffers from some long-lasting conceptual problems, namely,
the fine-tuning problem~\cite{Adams:1990pn}, the initial condition
problem~\cite{Ijjas:2013vea}, the trans-Planckian
problem~\cite{Martin:2000xs}, the measure problem~\cite{Gibbons:2006pa},
singularity problems~\cite{Borde:1993xh}, etc, which makes inflation
to be known as a paradigm rather than a well-established theory after three
decades.
 
Cosmologists believe that some, if not all, of the above problems might be solved by
consistently embedding an inflation model, taken as an effective field theory of gravity, 
into a M/string theory, taken as a candidate for a quantum theory of gravity. 
If this construction is possible, then one could in principle
build an ultraviolet (UV) complete model for inflation. Although
great efforts have been made to accomplish such a goal,
there is still no conclusive result for such possibility so far
(see, e.g., discussions in Ref.~\cite{Kachru:2003sx}). 
The most recent efforts led
into the so-called swampland conjectures~\cite{swampland conjectures},
which requests steep potentials $V$, such that
$M_{\rm Pl}\left|V^{\prime}\right|/V \gtrsim \mathcal{O}(1)$, with
a sub-Planckian field excursion, $\Delta \phi \lesssim M_{\rm Pl}$ (for a
general review on the swampland conjectures, see, e.g., Ref.~\cite{SCR} and
also the references therein for the many implications in cosmology). 
While the latter condition can be satisfied with
some inflationary models~\cite{Kallosh:2013hoa}, the former condition
given on the  potential rules out most slow-roll single field
inflationary models, requesting the violation of the slow-roll
regime. 

Among the many recent discussions concerning these conjectures, the authors
in Ref.~\cite{Kim:2019vuc} have
very recently explained the necessity of the swampland constraints
utilizing the completeness of the spectrum of charged branes  in a quantum
theory of gravity and the consistency of what lives on the branes. In
fact, they have shown that coupling branes, as a consistent ingredient
of higher dimensional theories, to quantum theories of gravity in
higher dimensions, are useful to separate theories which are in the
landscape from those which belong to the swampland. 
Baneworld type of models have been extensively used in cosmology.
In particular the braneworld models from Randall and Sundrum~\cite{Randall-Sundrum},
which we will focus in the present work, also known as Randall-Sundrum (RS) type I and II
models. In such theories, all the standard particles and
their interactions are confined to the brane and just gravity can
propagate along a fifth dimension. The extra dimension modifies the
{}Friedmann equation on the brane, leading in particular to a
quadratic term proportional to the energy density, which have some
significant cosmological implications~\cite{BraneCosmo}. 
The scenario of inflation in RS II braneworld was
studied in Ref.~\cite{Maartens:1999hf}, in which the authors have
shown that brane corrections at the high energy regime allow steep
potentials to be embedded in the braneworld scenario. 
More recently, the authors in Ref.~\cite{Cosmological implications} have also
discussed about the swampland in this model.

In the present work, we will focus on an inflation model with an
exponential potential in the context of the RS II braneworld
scenario. Let us recall that  exponential potentials for scalar fields
can naturally emerge in M/string theory from the compactification of
the higher dimensions into  a four-dimensional effective field theory,
that encodes the size of the extra dimensions~\cite{String}. It has
been shown that an exponential potential form for the scalar inflaton
field (and with canonical kinetic term) leads to a power-law type of
solutions, with the scale factor satisfying $a(t) = t^{q}$ and
$q>1$. This type of potential can also describe different cosmological
epochs, including, for instance, the behavior of the late-time
Universe~\cite{Copeland:1997et}. Moreover, it leads to an everlasting power
law inflation at the very early Universe, with constant slow-roll
parameters~\cite{Power-law, Yokoyama:1987an}.  Despite all of this, several attempts
have been made towards overcoming such difficulties, like, for instance,
implementing these exponential potential models in the braneworld
scenario~\cite{2A,Sahni:2001qp,Dias:2010rg}. Although inflation with an exponential
potential in the braneworld scenario can have a graceful exit, it is
still in tension with the observational data.
Even in a braneworld scenario, an exponential potential model for
inflation produces a large tensor-to-scalar ratio, so it is considered
excluded by the observational data~\cite{Dias:2010rg}.
It also requires an efficient reheating mechanism~\cite{2A}. Besides, it suffers from a
long kination period after inflation, leading to a large amount of
gravitational particle production and relic gravity waves, disrupting
and even undermining the big bang nucleosynthesis (BBN)~\cite{Sahni:2001qp}. 

Although dissipation is an indispensable part of any physical system
interacting with its environment, conventional inflationary scenarios,
namely cold inflation, typically neglect such effects during the
inflationary phase. Thus, a (p)reheating phase at the end of inflation
is required so to connect with the big bang radiation dominated
regime.  In the warm inflation (WI) picture~\cite{Warm inflation} the
possible dissipative effects that results from the interactions of the
inflaton field with other field degrees of freedom can sustain a
subdominant radiation bath throughout the inflationary dynamics. 
If the dissipative effects can become strong
enough, the Universe can smoothly enter into the radiation dominated era,
with no separate reheating phase being in general
required~\cite{Berera:1996fm}. 
There is by now an extensive literature on different aspects of WI,
the models giving origin to it and on its applications~\cite{Berera:1998hv,WIB,WDM,WDE1,WDE2,Lima:2019yyv,Steeper1,Steeper2,Perturbations1, Perturbations2, Benetti:2016jhf,Lower,Dymnikova:1998ps,Herrera:2018wan,Bastero-Gil:2016qru,WIS,Bastero-Gil:2019gao,WBI,Bastero-Gil:2019gao}. 
{}For instance, the inflaton itself can be a source and
responsible for cosmic magnetic field generation~\cite{Berera:1998hv}.
In combination with the intrinsic dissipative effects lead to a
successful baryogenesis scenario~\cite{WIB}. There can also be a
remnant inflaton field behaving like dark matter~\cite{WDM} and dark
energy~\cite{WDE1, WDE2, Lima:2019yyv}. {}Furthermore, such dissipation
effects change the dynamics of inflation due to the supplementary
friction term, making it possible to construct models with steep
potentials~\cite{Steeper1, Steeper2}. Moreover, the dissipation and
radiation effects modify the primordial spectrum of curvature
perturbations~\cite{Perturbations1, Perturbations2, Benetti:2016jhf},
resulting in a lower energy scale of inflation and making it also possible to
reconcile steep potentials with the observational data~\cite{Lower,
  Bastero-Gil:2016qru}. Such appealing features of WI allow it to
simultaneously satisfy the so-called swampland conjectures, provided
WI can occur with a sufficient strong
dissipation~\cite{WIS,Bastero-Gil:2019gao}. {}Furthermore, strong dissipation effects are
also able to suppress the energy scale of brane inflation, making it
consistent with the observations even in the high energy
regime~\cite{WBI}. Although it is enormously challenging achieving
strong dissipative regimes in WI, a recently proposed new model was
built using first principles in quantum field theory and shown
explicitly to overcome such difficulties~\cite{Bastero-Gil:2019gao}.  

It is the purpose of the present work to investigate WI in a model
with an exponential potential in the RS II braneworld scenario. We
are going to explicitly see how it is possible to make this WI model
realization fully functional and consistent with both observations and
also evading the theoretical restrictions faced by this type of model
in its previous constructions done in the context of cold inflation.
As an additional welcome feature, we also show that it can support
strong enough dissipation to evade the previous, but also the more
recent, swampland conjectures. 

The paper is organized as follows. In Sec.~\ref{sec2}, we briefly
review the WI dynamics constructed in the RS II braneworld
scenario. In Sec.~\ref{sec3}, we discuss several limitations imposed
on the exponential potential considered in this work. We also give the
motivations for considering such a potential and discuss how WI is
going to remove the discussed previous limitations imposed on the model. 
In Sec.~\ref{sec4}, we give the
explicit implementation of the model in WI and demonstrate
its viability through explicit numerical analysis. {}Finally, in
Sec.~\ref{concl}, we give our conclusions and also discuss the
viability of the model to unify the cosmology of the very early
Universe with the present epoch through a possible
quintessential inflation scenario.

\section{Warm inflation dynamics in the braneworld scenario}
\label{sec2}

In this section, we consider the RS II braneworld scenario in which
our four-dimensional world has its space dimensions like a 3 brane embedded 
in a five-dimensional bulk. In such a braneworld scenario, adopting a
{}Friedmann-Lema\^{i}tre-Robertson-Walker (FLRW) metric, the
cosmological evolution can be described by a generalized {}Friedmann
equation as follows~\cite{Shiromizu:1999wj, Maartens:1999hf}
\begin{equation}
H^2 \equiv \left( \frac{\dot a}{a} \right)^2 =\frac{1}{3 M_{\rm Pl}^2}
\rho + \frac{1}{36 M_5^6} \rho^2 + \frac{\Lambda_4}{3} +
\frac{\mu}{a^4},
\label{Hbrane}
\end{equation}
where $a(t)$ is the usual scale factor, $\rho$ is the total energy
density in the three-dimensional space,  $M_{\rm Pl} \equiv 1/\sqrt{8
  \pi G} \simeq 2.4 \times 10^{18}$GeV is the reduced Planck mass
defined in terms of the usual Newton gravitational constant $G$,
$\Lambda_4$ is an effective cosmological constant, $\mu$ is an
integration constant which behaves as dark radiation term, and  $M_5$
is the five-dimensional Planck mass, defined as
\begin{equation} \label{a1}
M_5^6 = \frac{\lambda M_{\rm Pl}^2}{6},
\end{equation}
where $\lambda$ is the intrinsic tension on the brane.  The
cosmological constant term in Eq.~(\ref{Hbrane}) can be tuned to zero
by an appropriate choice of the brane tension and bulk cosmological
constant. Also, the dark radiation term quickly redshifts during
inflation so we can also neglect it. Thus, Eq.~(\ref{Hbrane}) can then
be expressed simply as
\begin{equation}
H^2 =\frac{1}{3 M_{\rm Pl}^2} \rho \left( 1 + \frac{\rho}{2 \lambda}
\right).
\label{Hubble}
\end{equation}

In the WI scenario, the total energy density is given by 
\begin{equation}
\rho = \frac{{\dot \phi}^2}{2} + V(\phi) + \rho_R,
\label{rho}
\end{equation}
which accounts for the radiation fluid, with energy density $\rho_R$,
and the scalar field (the inflaton) $\phi$, with potential $V(\phi)$.
The background dynamics for the inflaton field $\phi$ and the
radiation energy density $\rho_R$, which are coupled to each other,
are, respectively, given by
\begin{eqnarray}
&& \ddot \phi + 3 (1+Q) H \dot \phi + V_{,\phi}=0,
\label{eqphi}
\\ && \dot \rho_R + 4 H \rho_R = 3H Q \dot \phi^2,
\label{eqrhoR}
\end{eqnarray}
where dots denote temporal derivatives and $Q$ is the dissipation
ratio in WI, defined as
\begin{equation}
Q= \frac{\Upsilon(T,\phi)}{3 H},
\label{Q}
\end{equation}
where $\Upsilon(T,\phi)$ is the dissipation coefficient in WI, which
can be a function of the temperature and/or the background inflaton
field, depending on the specifics of the microscopic physics behind
the construction of WI.  {}For a radiation bath of relativistic
particles, the radiation energy density is given by $\rho_R=\pi^2 g_*
T^4/30$, where $g_*$ is the effective number of light degrees of
freedom  ($g_*$ is fixed according to  the dissipation regime and
interactions form used in WI).  

A typical dissipation coefficient functional form in terms of the
temperature and the inflaton field amplitude found in many WI models
can be expressed as~\cite{Steeper1, Steeper2, BasteroGil:2010pb,
  Bartrum:2013fiaBerera:2018tfcGraef:2018ulgBastero-Gil:2018uep},
\begin{equation}
\Upsilon(T,\phi) = C T^c \phi^p M^{1-c-p},
\label{Upsilon}
\end{equation}
where $C$ is a dimensionless constant (that carries the details of the
microscopic model used to derive the dissipation coefficient, e.g.,
the different coupling constants of the model), $M$ is a mass scale in
the model and depends of its construction, while  $c$ and $p$ are
numerical powers, which can be either positive or negative
numbers (the dimensionality of the dissipation coefficient in
  Eq.~(\ref{Upsilon}) is of course preserved, i.e., $[\Upsilon] =
  [{\rm energy}]$).
Some typical dependencies found in the
literature, are the cases with $c=-1,\, p=2$ (see
Ref.~\cite{Gleiser:1993ea}),  $c=1,\, p=0$ (see
Ref.~\cite{Bastero-Gil:2016qru}), $c=-1,\,p=0$ (in the asymptotic high
temperature limit for the dissipation coefficient found in the model
of Ref.~\cite{Bastero-Gil:2019gao}) and $c=3,\, p=-2$ (see
Refs.~\cite{Steeper1, BasteroGil:2010pb}). The case with $c=3,\, p=-2$
was in fact the first microscopic construction for WI from a particle
physics model building perspective and giving consistent
observables~\cite{Steeper2}. In the present work we will not be interested
in the specific origin of these dissipation coefficients from a microscopic
quantum field theory derivation (for such details we refer the interested 
reader to the above cited references), but on the phenomenological consequences
of them. 

In terms of Eqs.~(\ref{Hubble}), (\ref{eqphi}) and (\ref{eqrhoR}) the
slow-roll parameters in the braneworld scenario are given by 
\begin{eqnarray}
\epsilon_{\rm brane} &=& \epsilon_V \frac{1+V/\lambda}{\left[1+V/(2
    \lambda)\right]^2},
\label{epsb}
\\ \eta_{\rm brane} &=& \eta_V \frac{1}{1+V/(2 \lambda)},
\label{etab}
\end{eqnarray}
where $\epsilon_V$ and $\eta_V$ are the usual slow-roll inflaton
potential parameters, defined as
\begin{eqnarray}
\epsilon_V &=& \frac{M_{\rm Pl}^2}{2} \left( \frac{V_{,\phi}}{V}
\right)^2,
\label{epsV}
\\ \eta_V &=& M_{\rm Pl}^2 \frac{V_{,\phi\phi}}{V}.
\label{etaV}
\end{eqnarray} 
In terms of Eqs.~(\ref{epsb}) and (\ref{etab}), the slow-roll
conditions in WI are  defined by the requirement that $\epsilon_{\rm
  brane} \ll 1 + Q$ and $\eta_{\rm brane} \ll 1+Q$.  The accelerated
inflationary dynamics terminates when $\epsilon_{\rm brane} = 1+Q$.

By assuming the slow-roll approximation, with energy density $\rho
\sim V$, the Eqs.~(\ref{Hubble}), (\ref{eqphi}) and (\ref{eqrhoR})
reduce to
\begin{eqnarray}
&&H^2 \simeq \frac{1}{3 M_{\rm Pl}^2} V \left( 1 + \frac{V}{2 \lambda}
  \right), \\ &&  3 (1+Q) H \dot \phi \simeq - V_{,\phi}, \\  &&  4
  \rho_R \simeq 3 Q \dot \phi^2.
\end{eqnarray}

Given the dissipation coefficient with the functional form given by
Eq.~(\ref{Upsilon}) and expressing the evolution in terms of the
number of e-folds, $dN = H dt$, we can deduce how the dissipation ratio
$Q$ and the ratio of temperature of the radiation bath by the Hubble
rate, $T/H$, evolve during WI. After some straightforward algebra, we
find that their evolution are determined by the equations
\begin{widetext}
\begin{eqnarray}
\frac{d \ln Q}{dN} &=& \frac{2\left[ (2 + c) \epsilon_{\rm brane} - c
    \, \eta_{\rm brane} -  2 p\, \kappa_{\rm brane} \right]}{4 - c +
  (4 + c) Q},
\label{dQdN}
\\ \frac{d\ln (T/H)}{dN} &=& \frac{ \left[7 + c \, (Q-1) + 5 Q\right]
  \epsilon_{\rm brane} -  2 (1 + Q) \eta_{\rm brane} + (Q-1)p\,
  \kappa_{\rm brane} } {(1 + Q) [4 - c + (4 + c) Q]},
\label{dTHdN}
\end{eqnarray}
\end{widetext}
where $\kappa_{\rm brane}$ is defined as
\begin{equation}
\kappa_{\rm brane} =  M_{\rm Pl}^2 \frac{V_{,\phi}}{\phi \, V \left[1
    + V/(2 \lambda)\right]}.
\label{kappa}
\end{equation}
Moreover, the ratio of radiation to inflaton energy density in the
slow-roll regime is roughly given by
\begin{align}
\frac{\rho_{R}}{{\rho_{\phi}}} \approx \frac{1}{2} \frac{\epsilon_{\rm
    brane}}{1+Q} \frac{Q}{1+Q}.
\end{align}
Although the ratio of the radiation to inflaton energy density is very
small at the beginning of WI, it can be  large at the end of inflation
(when $\epsilon_{\rm brane} = 1+Q$), even for an initially small dissipation ratio
$Q$.  Consequently, the Universe can smoothly enter into the radiation
dominated epoch at the end of WI,  with no need for a separate
reheating phase a \textit{priori}.

\section{The model}
\label{sec3}

As explained in the Introduction, in this paper we will be working
with the exponential potential form for the inflaton field, 
\begin{equation}
V(\phi) = V_0 \exp\left(-\alpha \phi/M_{\rm Pl}\right),
\label{Vphi}
\end{equation}
where $V_0$ is the normalization for the potential and which can be
fixed by the amplitude of primordial curvature perturbations as usual.
The study of inflation with this potential is diverse, specially
motivated  by the fact that a potential like Eq.~(\ref{Vphi}) can also
serve as a quintessential inflation model, describing both the
inflationary early Universe period, as also the late Universe, working
as an evolving dark energy model at the present
epoch~\cite{2A}.  The exponential potential for the
inflaton has also already been studied before in the presence  of
dissipative effects from particle production~\cite{Yokoyama:1987an},
yet with no connection to the WI scenario, as it is the focus in the present
study.

Despite the many uses of the potential Eq.~(\ref{Vphi}), it is,
however, hard to find consistent  ways of ending inflation in such
models in the traditional ways (i.e., ending the acceleration regime).
{}For instance, with the standard slow-roll parameters given by
Eqs.~(\ref{epsV}) and (\ref{etaV}), the accelerated expansion requires
$\alpha < \sqrt{2}$. However, this leads to a power-law type of
inflation, which has long been found to be inconsistent with the
observations~\cite{Hinshaw:2012aka, Planck:2013jfk}, unless there are
either modifications to the form of the potential, to the dynamics, or
to both.  In the braneworld scenario, it is possible to have inflation
even when $\alpha > \sqrt{2}$, provided that the dynamics is dominated
by the high energy brane regime, $V > 2 \lambda$. In this case, the
modification of the dynamics because of the brane corrections are
strong and we also have a stronger Hubble friction. As the inflaton
field evolves and $V < 2 \lambda$, the effects of the brane
corrections lessen, the usual general relativity (GR) dynamics is
restored, and inflation ends. But the  inflaton evolution in a steep
potential like Eq.~(\ref{Vphi}) will at some point be dominated by the
kinetic energy of the field, which gives start to a kination
regime. During this regime, the energy density  falls off like stiff
matter, $\rho \propto 1/a^6$. The abrupt change in the dynamics  can
produce a large amount of gravitational particle production that might
disrupt BBN later on~\cite{2A}. {}Furthermore, a large
amount of relic gravity waves is also  predicted to be
generated~\cite{Sahni:2001qp}. The energy density on these produced
gravitational waves can be boosted by the kination period and can also
disrupt the BBN later on. As an additional
problem, the model leads yet to predictions on the tensor-to-scalar
ratio and on the spectral tilt of the primordial scalar spectrum 
that is also excluded by the observations~\cite{Dias:2010rg}.  All
these issues together have rendered this inflaton potential model
unsuitable from an observational point of view. 

In the following, we discuss the choice for the dissipation
coefficient used in the present study. We will also discuss the
theoretical and observational constraints on the parameters of the
braneworld scenario and on the exponential potential.  Then, we
explain under which conditions the WI scenario realization of the
model may allow it to overcome all of the above mentioned difficulties,
and we then give an explicit example in Sec.~\ref{sec4}. 

\subsection{Warm inflation implementation}

Considering the model given by  Eq.~(\ref{Vphi}) in the context of WI,
we find, in particular, that the evolution equation for the dissipation
ratio Eq.~(\ref{dQdN}) becomes
\begin{widetext}
\begin{align}
\frac{d \ln Q}{dN}= \frac{4 \lambda \alpha \left\{ 2 p (2 \lambda + V)
  + \left[(2 - c) \lambda + 2 V\right] \alpha \phi/M_{\rm Pl} \right\}
}{\left[4 - c + (4 + c) Q\right] (2 \lambda + V)^2 \phi/M_{\rm Pl} }.
\label{dQdNexp}
\end{align}
\end{widetext}
We are interested in the strong dissipative regime of WI, i.e., $Q \gg
1$, which ensures that the swampland conditions are
satisfied~\cite{WIS}, with the inflaton  excursion typically
sub-Planckian, $\Delta \phi < M_{\rm Pl}$. {}Furthermore, we have the
constraint to be discussed below and given by Eq.~(\ref{alphabound}), 
on the inflaton potential constant $\alpha$, requiring rather large values.  
At the same time, we do not want the dissipation
ratio to grow throughout the evolution; otherwise, even by arranging
inflation to end due to the braneworld effects, it might reinitiate
again or even never end in the first place. Hence, we would ideally
want a {\it decreasing} $Q$ with the number of e-folds, i.e., the
right-hand side of Eq.~(\ref{dQdNexp}) should become  negative, soon
after the braneworld effects become subdominant.  Thus, we need to
analyze the region of parameters for which the right-hand side of
Eq.~(\ref{dQdNexp}) can be either  positive or negative.  In
particular, from previous studies on the stability of the WI
dynamics~\cite{Moss:2008yb}, we have
that the power $c$ should satisfy  $-4 < c < 4$, which results in the
denominator of Eq.~(\ref{dQdNexp}) to be always positive.  Thus, for
$2 < c < 4$ and $-3<p<0$, we find that the numerator in the
right-hand side of Eq.~(\ref{dQdNexp}) can be positive in the high
energy regime, $V \gg 2\lambda$, but then it becomes negative as one
enters in the low energy regime, $ V < 2 \lambda$. This is, in
particular, exactly the behavior expected with a dissipation
coefficient with a functional form with a cubic power in the
temperature, as found, e.g., in Refs.~\cite{Steeper1,
  BasteroGil:2010pb}. In this case, we find that $Q$ will grow
initially during the inflationary evolution, provided that $V> 2
\lambda$, decreasing  later on in the evolution when $V < 2
\lambda$. Thus, by having initially that $\epsilon_{\rm brane}/(1+Q) \ll 
1$,  this will ensure that the accelerated expansion will happen both
because of the brane effect and because of a growing $Q$, further
facilitating the accelerated expansion, even if $\alpha$ is large.  On
the other hand, as the brane effects lessen and we recover the usual
general relativity evolution, the dissipation ratio will start to
decrease, eventually ending the accelerated inflationary regime due to
the large value of $\alpha$. Note that this effect of the dissipation
on the evolution does not happen when $c=1$, as in the case of the
dissipation coefficient found  in Ref.~\cite{Bastero-Gil:2016qru}, or
in the case of the dissipation coefficient recently found and
correspondingly, WI dynamics studied  in
Ref.~\cite{Bastero-Gil:2019gao}, which favors a dissipation
coefficient with $c=-1$. In both of these two cases, $Q$ will always
increase, in general, making it more difficult to end inflation. In fact,
such decreasing $Q$ allowed by a $\Upsilon \propto T^3/\phi^2$
dissipation  coefficient not only plays a pivotal role in bringing the
inflation to an end but also it will be shown to allow the model
studied here to be consistent with the observations, even for large
values of the dissipation ratio $Q$.  Thus, in the following, we will restrict
our study to the cubic in the temperature dissipation coefficient
form, where\footnote{The attentive reader may question what happens
  with this dissipation coefficient with the exponential potential we
  use here and where $\phi$ might cross zero. This is not a problem
  since, as shown in  Refs.~\cite{Steeper1, Steeper2,
    BasteroGil:2010pb,
    Bartrum:2013fiaBerera:2018tfcGraef:2018ulgBastero-Gil:2018uep},
  this dissipation coefficient is derived when the inflaton field is
  coupled to heavy intermediate fields $\chi$, where the mass $m_\chi
  \propto \phi$, but we could as well have a nonvanishing bare mass
  for $\chi$, which regulates any possible infrared divergence and
  avoids $\Upsilon$ to diverge when $\phi \to 0$.}
\begin{equation}
\Upsilon(T,\phi) = C \frac{T^3}{\phi^2}.
\label{cubic}
\end{equation}

In the next section, we will then show our explicit and detailed
results for the WI model with the dissipation coefficient
Eq.~(\ref{cubic}) in the context of the braneworld scenario. But
before entering in the explicit results derived from this model, let
us discuss some important constraints that the model might be
subjected.

\subsection{Constraints}

In the braneworld scenario, the quadratic term of the energy density
in the Hubble parameter Eq.~(\ref{Hubble}) becomes dominant in the
high energy regime, provided the brane tension $\lambda$ has a moderate value. 
However, BBN implies that it has to be subdominant, since it
decays as $a^{-8}$ and becomes rapidly negligible
thereafter. Therefore, the BBN bounds put a lower limit
on the brane tension such that it has to satisfy $\lambda\gtrsim (1 {\rm MeV})^{4}$, 
which in combination with Eq.~(\ref{a1}) gives~\cite{Maartens:1999hf}
\begin{align}
M_{5}\gtrsim \left(\frac{1 \,{\rm MeV}}{M_{\rm Pl}}\right)^{2/3} \sim
10 \, {\rm TeV}.
\end{align}
However, considering the fifth dimension to be infinite and requiring
relative corrections to the Newtonian law of gravity to be also
small~\cite{Hoyle:2000cv}, one obtains a more stringent constraint as
given by $M_{5}\gtrsim 10^{5}\, {\rm TeV}$, or, equivalently,
$\lambda\gtrsim 100 \, {\rm GeV}$.

As discussed in Refs.~\cite{Kawasaki:2004yh,Copeland:2005qe}, the
abundance of gravitinos is related to the reheating temperature
through the Boltzmann equation. Hence, by constraining the gravitinos
abundance, one can obtain an upper bound on the reheating temperature as
$T_R \lesssim 10^6$--$10^8$ GeV. However, the authors in
Ref.~\cite{Copeland:2005qe} have shown that the relation between
gravitinos abundance and the reheating temperature breaks in the high
energy limit, $V/(2\lambda)>1$, and allows inflation to occur even
with a higher reheating temperature. 

The late-time behavior for steep exponential potentials
produces a scaling solution, where the scalar field  exhibits the same
redshift dependence as the dominant fluid in the Universe. Thus, the
energy density fraction for the scalar field, assuming spatial flatness,
has a scaling solution
given by~\cite{Copeland:1997et,2A}
\begin{equation}
\Omega_\phi = \frac{3(w+1)}{\alpha^2},
\label{Omegaphi}
\end{equation}
where $w$ is the equation of state of the dominant fluid. If the
inflaton potential Eq.~(\ref{Vphi}) remains unchanged till late in the
Universe, it will act like a quintessence field. Early dark energy can
influence the cosmic microwave background (CMB) peaks among to other
changes in the CMB power spectrum. Hence, its fraction can be strongly
constrained when including small-scale measurements and CMB lensing, for instance.  The strongest constraint on the fraction of dark energy
at last scattering time has been produced by the  Planck data and
leading to an upper bound~\cite{Ade:2015rim}, $\Omega_\phi \lesssim
0.0036$ at 95$\%$ confidence level (for Planck
TT,TE,EE-lowP+BSH). Applying this to Eq.~(\ref{Omegaphi}), we are lead
to the lower bound on the coefficient $\alpha$ for the inflaton
potential,
\begin{equation}
\alpha \gtrsim 33.3.
\label{alphabound}
\end{equation}

{}Furthermore, the recently proposed swampland conjectures restrict
both the dynamics and potential form of the inflationary models~\cite{swampland conjectures,SCR}. 
In fact, the de Sitter swampland conjecture requests steep potentials,
$M_{\rm Pl} V^{\prime}/V\ge \mathcal{O}(1)$, with no extrema,
as a result of which not only inflationary models but also the usual
reheating mechanism, due to an oscillatory phase around the minimum at the end
of inflation, are ruled out~\cite{Kamali:2019hgv}. Moreover, the
distance swampland conjecture requests sub-Planckian field excursions
during inflation, $\Delta \phi \lesssim M_{\rm Pl}$, 
whereby all large field inflationary models are
excluded when the conjecture is taken at face value. As discussed in
Ref.~\cite {WIS}, to have a large slow-roll parameter $\epsilon_{V}$,
one needs to violate the relation between $\epsilon_{H}\equiv
-\dot{H}/H^2$ and $\epsilon_{V}$ in the conventional single field
slow-roll inflation. Besides, one also needs to find another mechanism
to heat up the Universe at the end of inflation. The WI scenario has
the appealing feature to give us both of these in a single frame due
to the presence of the intrinsic dissipation effects, provided that WI
occurs in the strong dissipative regime~\cite {WIS}, $Q\gg1$. 
Moreover, the WI scenario also gives us the possibility of suppressing the
tensor-to-scalar ratio by several orders of magnitude for a large
dissipation ratio, due to effects of the dissipation on the
primordial perturbations and the change in the scalar curvature
spectrum (to be discussed below). As a result, the inflaton field can
remain sub-Plankian during inflation, evading the Lyth bound even for
steep potentials. Therefore, dissipation effects make WI able to
satisfy the swampland conjectures, while also allowing for the accelerated expansion 
even when $\alpha>1$ in the exponential potential and braneworld scenario as
considered here.

In our numerical study to be presented in the next section,  we will
keep in mind all of the above constraints and conditions and aim at
satisfying them all.

\section{Numerical results}
\label{sec4}

As previously mentioned, both the background dynamics and the
perturbations get modified due to the presence of dissipation and a
radiation bath during WI~\cite{Perturbations1, Perturbations2}. {}As
a matter of fact, the primordial power  spectrum for WI at horizon
crossing can be expressed in the form (see, e.g.,
Ref.~\cite{Benetti:2016jhf} and references therein),
\begin{equation} \label{Pk}
\Delta_{{\cal R}}(k/k_*) =  \left(\frac{ H_{*}^2}{2
  \pi\dot{\phi}_*}\right)^2  {\cal F} (k/k_*),
\end{equation}
where the subindex ``$*$" stands for those quantities evaluated at the
Hubble radius crossing,  $k_*=a_* H_*$.  The function ${\cal F}
(k/k_*)$ in Eq.~(\ref{Pk})  is given by
\begin{equation}
{\cal F} (k/k_*) \equiv  \left(1+2n_* + \frac{2\sqrt{3}\pi
  Q_*}{\sqrt{3+4\pi Q_*}}{T_*\over H_*}\right) G(Q_*),
\label{calF}
\end{equation}
where $n_*$ denotes the inflaton statistical distribution due to the
presence of the radiation bath and $G(Q_*)$ accounts for the effect of
the coupling of the inflaton fluctuations to
radiation~\cite{Perturbations2}.  $G(Q_*)$, in general, can only be
determined by numerically solving the full set of perturbation equations in
WI and fitting it to an appropriate function. {}Following an analogous 
derivation as considered, e.g., in the
papers in Ref.~\cite{Perturbations2}, we find that an appropriate
functional form for $G(Q_{\star})$ that is valid for the  present
exponential model in the braneworld construction and with dissipation
coefficient (\ref{cubic}),  is well described by 
\begin{eqnarray}
G(Q_*) &=& \frac{1 + 0.413 Q_*^{0.85}}{\left(1 + 0.18 Q_*^{0.859}
  \right)^{26.5}}  \nonumber \\ &+&  \frac{0.00692 \exp\left(6.44
  Q_*^{0.292}\right)}{1 + 0.00082 \exp\left(0.1763
  Q_*^{0.66}\right)}.
\label{GQ}
\end{eqnarray}
The above equation for $G(Q_*) $ is found to hold for rather very
large values for $Q_*$, up to around  $Q_* \simeq 2000$. Besides,
since the behavior of the spectrum with $Q_*$ is smooth and well
behaved, we can always do this procedure (numerical fitting) with a
sufficient precision such that any arbitrariness in the numerical
fitting does not change the observable quantities, e.g., the spectral
index $n_s$ and the tensor-to-scalar ratio $r$, both defined below.

In our numerical results, we fix the scalar spectral amplitude value at
the pivot scale $k_*$  as  $\ln\left(10^{10} \Delta_{{\cal R}} \right)
\simeq 3.047$, according to the Planck
Collaboration~\cite{Aghanim:2018eyx} (in the TT,TE,EE-lowE+lensing+BAO
68$\%$ limits data set).

While the primordial scalar curvature perturbation in WI gets modified
according to Eq.~(\ref{Pk}), the tensor perturbations spectrum  is
unchanged in WI because of the weakness of the gravitational
interactions~\footnote{See, however, Ref.~\cite{Li:2018wno} for possible
changes in the tensor spectrum due to WI. However, even for the very large
values of $Q_*$ considered here by us, those corrections found 
in Ref.~\cite{Li:2018wno} are completely negligible and can be
safely neglected.}.  However, the spectrum of tensor perturbations is
modified in the braneworld scenario due to the presence of the extra
dimension, where the graviton resides, with respect to the standard
four-dimensional (GR) Universe.  The tensor perturbations power spectrum in
the braneworld scenario has been determined to be given
by~\cite{Langlois:2000ns}
\begin{equation}
\Delta_{T} = \frac{2 H^2}{\pi^2 M_{\rm Pl}^2} F^2(x),
\label{tensor}
\end{equation}
where
\begin{equation}
{}F(x) =  \left[ \sqrt{1+x^2} -x^2 \ln \left( \frac{1}{x} + \sqrt{ 1+
    \frac{1}{x^2}}\right)\right]^{-1/2},
\label{Fx}
\end{equation}
with $x=\sqrt{6 M_{\rm Pl}^2 H^2 /\lambda}$. 

{}From Eqs.~(\ref{Pk}) and (\ref{tensor}), the tensor-to-scalar ratio
$r$ is defined as usual,
\begin{equation}
r= \frac{\Delta_{T}}{\Delta_{{\cal R}}},
\label{eq:r}
\end{equation}
while the spectral tilt $n_s$ is defined as
\begin{equation}
n_s -1 = \lim_{k\to k_*}   \frac{d \ln \Delta_{{\cal R}}(k/k_*) }{d
  \ln(k/k_*) }.
\label{eq:n}
\end{equation}
We recall that the recent data from the Planck
Collaboration~\cite{Akrami:2018odb} has placed the upper bound
$r<0.056$ (95$\%$ C.L., Planck TT,TE,EE+lowE+lensing+BK15, at the pivot
scale $k_p = 0.002/{\rm Mpc}$), while for the spectral tilt the result
is $n_s=0.9658\pm 0.0040$  (95$\%$ C.L., Planck
TT,TE,EE+lowE+lensing+BK15+BAO+running).

In the results shown below, the number of e-folds for inflation, in
particular, the number of e-folds before Hubble radius crossing, $N_*$,
is always computed self-consistently.  This is done by noticing that
length scales crossing the Hubble radius during inflation and
reentering today will satisfy $k_*=a_i H_i = a_0 H_0$, such
that~\cite{Liddle:2003as}
\begin{equation}
\frac{k_*}{a_0 H_0} = e^{-N_*} \frac{T_0}{T_{\rm end}}
\frac{H_i}{H_0},
\label{N*}
\end{equation}
where $0$-index quantities mean that they are evaluated today, while $i$-index quantities
are those evaluated at $N_*$ e-folds before the end of inflation.  The CMB
temperature today, $T_0$, is set to the value $T_0 = 2.725\, {\rm K}=
2.349 \times 10^{-13}\, {\rm GeV}$.  In our convention, we use $a_0=1$
and for the Hubble parameter today,  we assume the Planck result,
$H_0=67.66\, {\rm km}\, s^{-1} {\rm Mpc}^{-1}$ [from the Planck
Collaboration~\cite{Aghanim:2018eyx}, TT,TE,EE-lowE+lensing+BAO 68$\%$
limits,  $H_0 = (67.66 \pm 0.42)\, {\rm km}\, s^{-1} {\rm Mpc}^{-1}$].
$T_{\rm end}$ is the temperature at the beginning of the radiation
dominated regime, which in the WI, turns out to be simply the
temperature at the end of inflation, since WI ends by the time the
radiation energy density takes over the inflaton one.  {}For the
exponential potential Eq.~(\ref{Vphi}) in the cold inflation case and
in the braneworld scenario, there is a prediction for $N_*$ given
by~\cite{Sahni:2001qp} $N_* \simeq 70$.  In the WI case studied here,
we always find a smaller value for $N_*$ due to the effect of
dissipation.

\begin{center}
\begin{figure}[!htb]
\includegraphics[width=8.6cm]{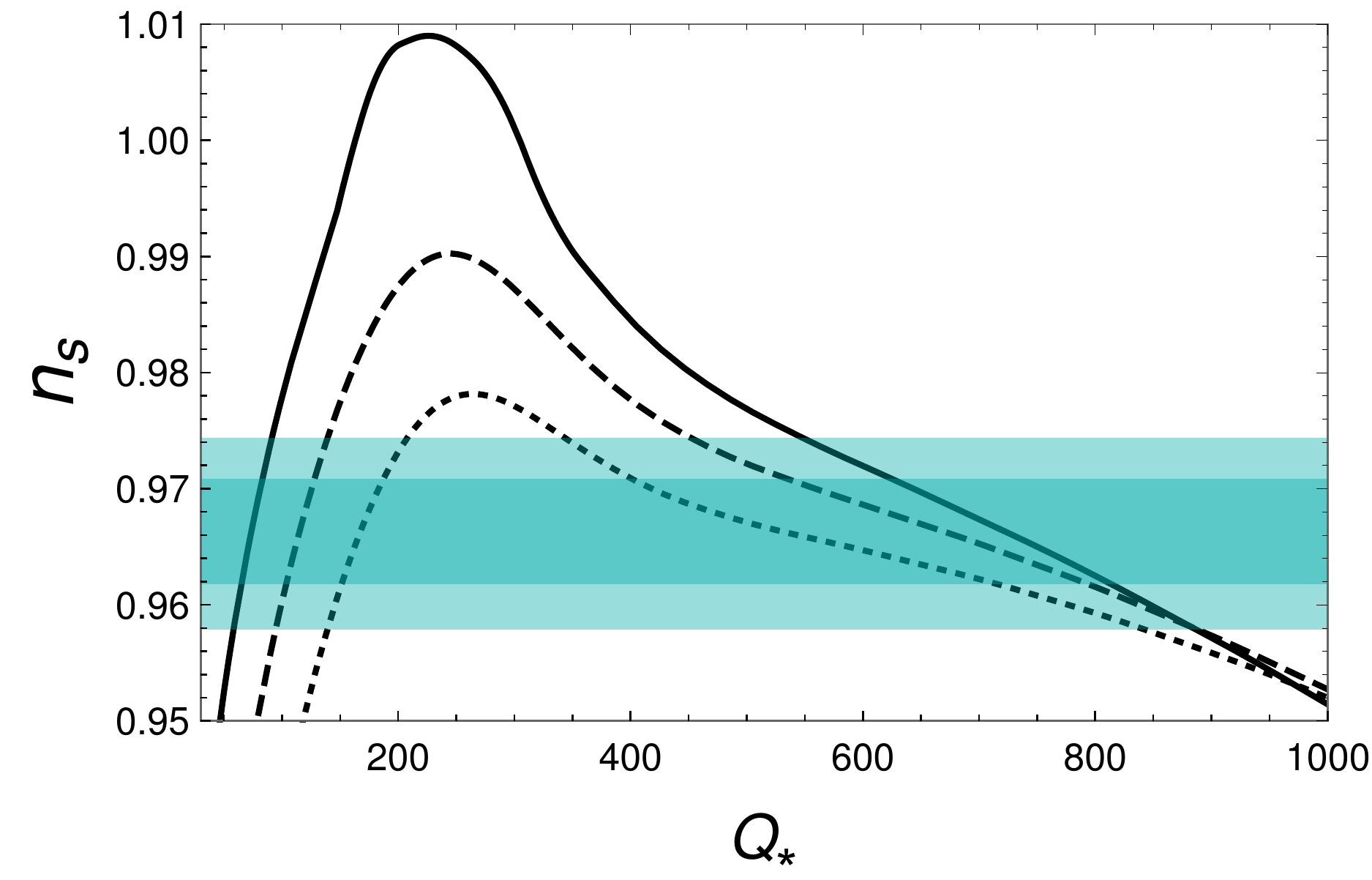}
\caption{The spectral index $n_s$ as a function of the dissipation
  ratio $Q_*$ for the cases of $\alpha=30$ (solid line), $\alpha=40$
  (dashed line), and $\alpha=50$ (dotted line), for a fixed value of
  (normalized) brane tension,  $\lambda/V_0= 5\times 10^{-5}$.  The
  shaded areas are for the $68\%$ and $95\%$ C.L. results from Planck
  2018 (TT+TE+EE+lowE+lensing+BK15+BAO data). }
\label{fignsXQlambda}
\end{figure}
\end{center}

In {}Fig.~\ref{fignsXQlambda}, we show the behavior of the spectral
index $n_s$ as a function of the dissipation ratio $Q_*$ at a Hubble
radius crossing, when keeping the ratio of the brane tension by the
normalization of the inflaton potential, $\lambda/V_0$, fixed but for
three different values for the exponent $\alpha$ in the inflaton
potential.  {}For a strong dissipation ratio, $Q_* \gg 1$, the
tensor-to-scalar ratio is always quite very small, $r<10^{-14}$. We
recall that such very small values for $r$ are typical for WI in the
strong dissipative regime. This is so because the scalar curvature
power spectrum Eq.~(\ref{Pk}) is completely dominated by the
dissipation, thus strongly suppressing the tensor-to-scalar
ratio. Whereby, when $Q_* \gg 1$, the inflaton statistical
distribution term $n_*$ in Eq.~(\ref{Pk}) also has a negligible effect, and we can set it as vanishing.  {}From {}Fig.~\ref{fignsXQlambda}, we
see that the larger is the $\alpha$, the steeper is the inflaton
potential, therefore, we can find a larger range of dissipation values
for which $n_s$ agrees with the observational data. We also see that
there are always two ranges of $Q_*$ values satisfying the
observations. 

\begin{center}
\begin{figure}[!htb]
\includegraphics[width=8.6cm]{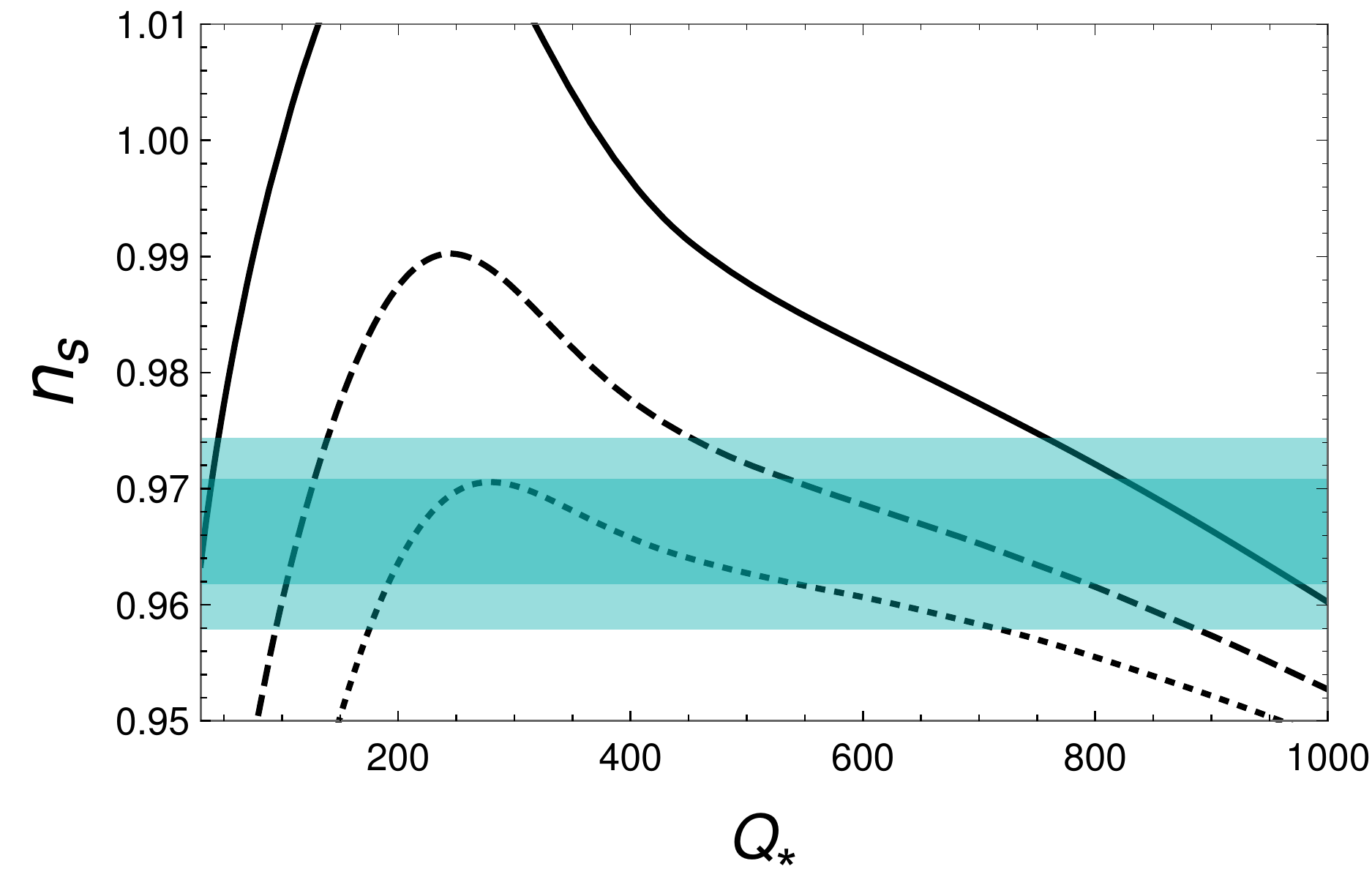}
\caption{The spectral index $n_s$ as a function of the dissipation
  ratio $Q_*$ for the cases of $\lambda/V_0= 10^{-5}$ (solid line),
  $\lambda/V_0= 5\times 10^{-5}$ (dashed line)  and $\lambda/V_0=
  10^{-4}$ (dotted line), for the fixed value of $\alpha=40$.  The
  shaded areas are for the $68\%$ and $95\%$ C.L. results from Planck
  2018 (TT+TE+EE+lowE+lensing+BK15+BAO data). }
\label{fignsXQalpha}
\end{figure}
\end{center}

In {}Fig.~\ref{fignsXQalpha}, we show the behavior of the spectral
index $n_s$ as a function of the dissipation ratio $Q_*$ at the Hubble
radius crossing, where now we keep the  constant $\alpha$
in the exponential inflaton potential fixed and give the results for three
different  values for  the ratio of the brane tension by the
normalization of the inflaton potential.  We have a similar behavior
as seen in {}Fig.~\ref{fignsXQlambda} when considering $\lambda/V_0$
fixed.  However, the larger is $\lambda/V_0$, the smaller is the range
of dissipation ratios satisfying the observations. In particular, for
$\alpha=40$, we have obtained that when  $\lambda/V_0 \gtrsim 10^{-3}$
there are no longer values for $n_s$ found to be compatible with the
Planck data, with $n_s$ being too red tilted.

\begin{center}
\begin{figure}[!htb]
\subfigure[ ]{\includegraphics[width=8.2cm]{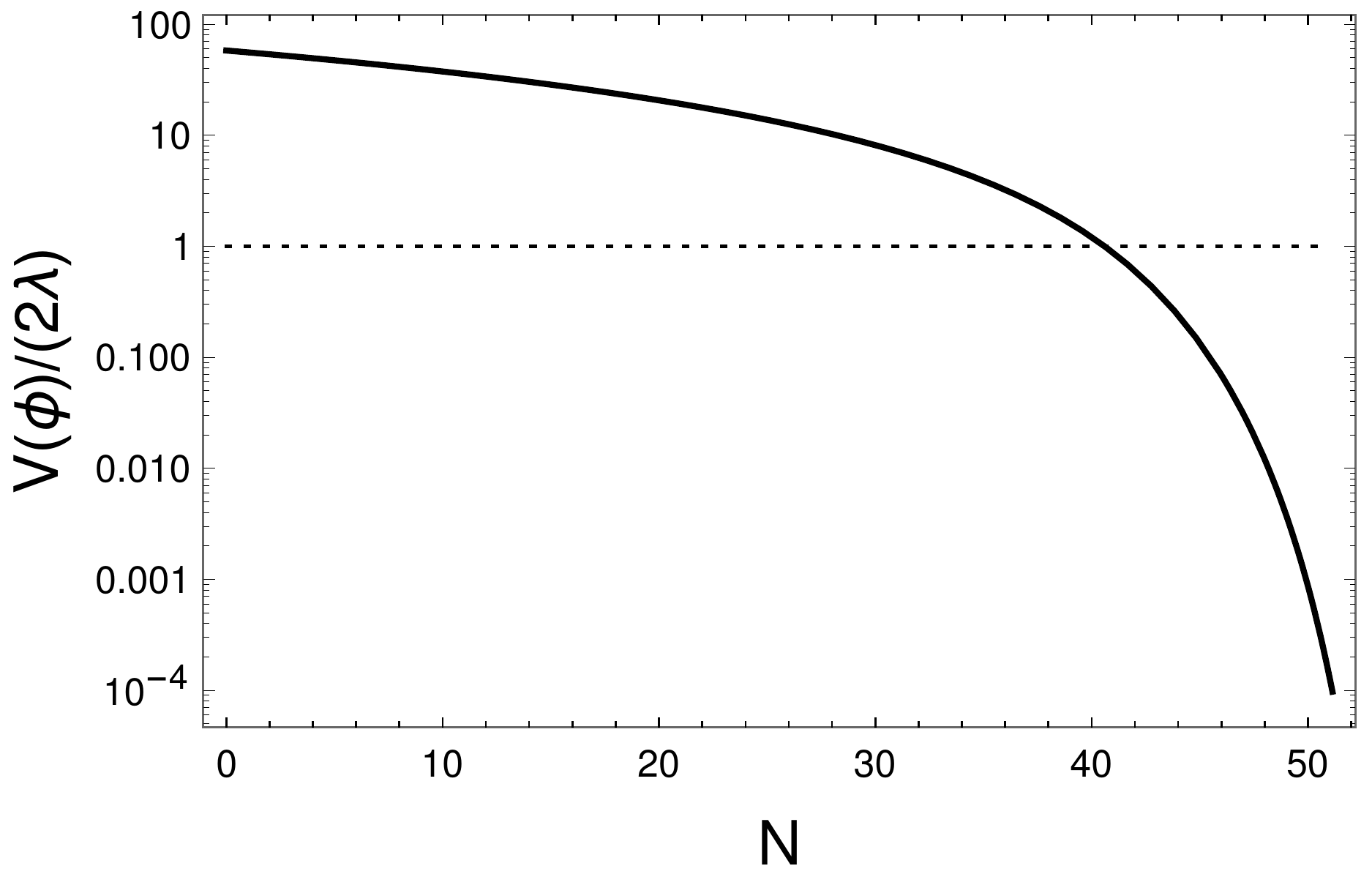}}
\subfigure[ ]{\includegraphics[width=8.2cm]{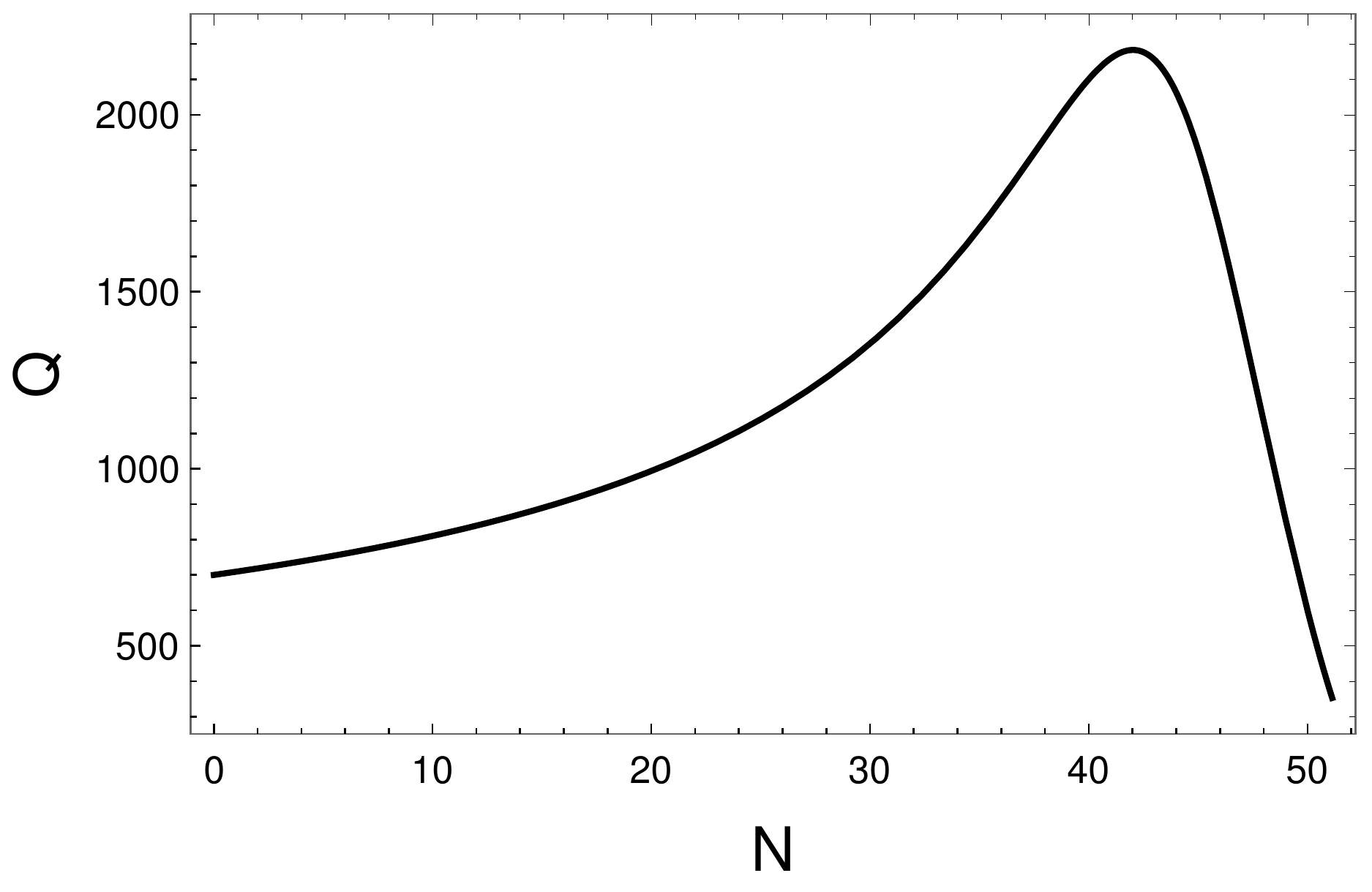}}
\caption{The evolution of the ratio $V(\phi)/(2 \lambda)$ [panel (a)]
  and the dissipation ratio $Q$ [panel (b)],  with the number of e-folds,
  for the choice of parameters $\alpha=40$ and $\lambda/V_0=5\times
  10^{-5}$. }
\label{figVQ}
\end{figure}
\end{center}

In {}Fig.~\ref{figVQ}(a), we show the evolution of the ratio $V(\phi)/(2
\lambda)$ with the number of e-folds, while in {}Fig.~\ref{figVQ}(b) we
show the evolution of the dissipation ratio with the number of e-folds.
We have chosen the particular case of $\alpha=40$ and
$\lambda/V_0=5\times 10^{-5}$, but there is little change in the
results when considering other values of parameters close to
these. This is found to be true for all the background quantities in
general.  {}From these results we can confirm the behavior for the
dissipation ratio with the cubic dissipation coefficient anticipated
in the previous section. In the high energy regime, $V > 2 \lambda$,
i.e., when the brane corrections prevail, $Q$ grows with the the
number of e-folds, while when the brane corrections lessen,  $V < 2
\lambda$, $Q$ fast decreases.

\begin{center}
\begin{figure}[!htb]
\includegraphics[width=8cm]{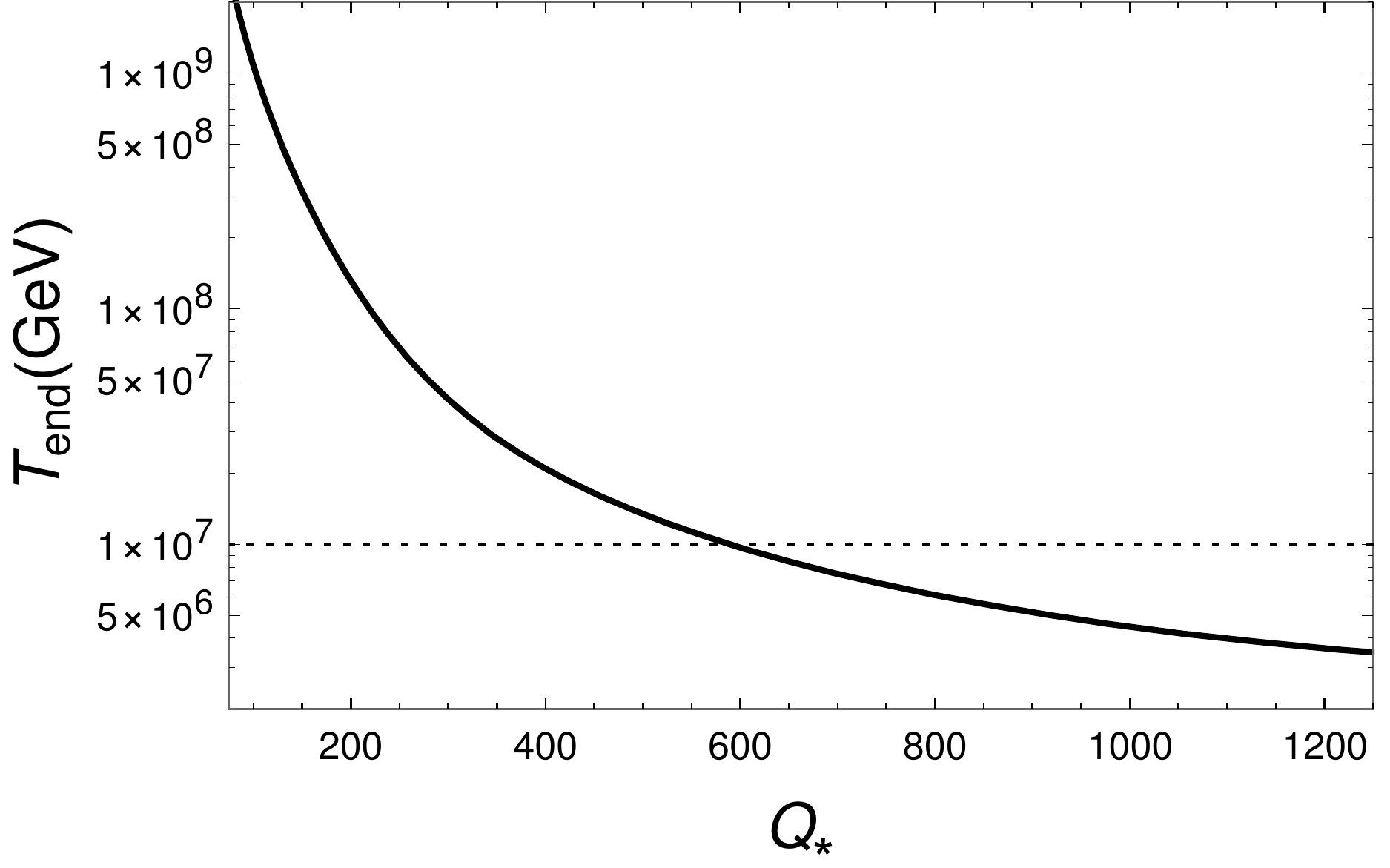}
\caption{The temperature at the end of inflation as a function of the
  dissipation ratio $Q_*$ for $\alpha=40$ and $\lambda/V_0=5\times
  10^{-5}$.}
\label{figTXQ}
\end{figure}
\end{center}

Considering the same parameters as those used in the previous figure,
in  {}Fig.~\ref{figTXQ} we show the behavior of the temperature at the
end of inflation as a function of the dissipation ratio.  The thin
horizontal dotted line at $T=10^{7}$GeV has been drawn as reference
only.  {}From the results shown in {}Fig.~\ref{figTXQ}, we see that the
gravitino bound can comfortably be satisfied for a sufficient large
dissipation and the result is also consistent with the Planck data, as
seen from {}Figs.~\ref{fignsXQlambda} and \ref{fignsXQalpha}.

We note that by taking as an explicit example for illustration 
purposes, the choice of parameters
$\alpha=40$, $\lambda/V_0=5\times 10^{-5}$, and $Q_*=700$, we find that
$n_s\simeq 0.965$; that is well within the central value obtained from the
Planck legacy data. {}Furthermore, as expected for such a large
dissipation ratio, we have an extremely small tensor-to-scalar ratio,
$r \simeq 1.3 \times 10^{-27}$.  We also find for these parameter values that
the inflaton potential normalization is ${V_0}^{\frac{1}{4}} \simeq
2.219 \times 10^9 {\rm GeV}$ and $N_* \simeq 51.1$.  {}For these
parameters, we also find that $V_*^{1/4}\simeq 6.12 \times 10^8$ GeV
and that at the end of the inflationary regime, $V_{\rm end} \simeq
2.20 \times 10^7$ GeV.  {}Furthermore, we find that $T_{\rm end}
\simeq 7 \times 10^6$GeV at the end of inflation.  The same parameters
produce a sub-Planckian inflaton field excursion, found to be  $\Delta
\phi \simeq 0.3\, M_{\rm Pl}$. In addition, the inflaton potential
slow-roll parameter, $\epsilon_V$, 
can be larger than one, as a consequence of the large value of
the dissipation coefficient, while still allowing accelerated 
expansion by
having $\epsilon_H \equiv -\dot{H}/H^2 \ll 1$.  
These results alone
already ensure that the distance and the de Sitter (along also with
the refined) swampland conjectures~\cite{swampland conjectures,WIS}
are satisfied.  In addition, given the lower energy scale for
inflation allowed for the present model in the large dissipation
regime, we also find the recently proposed trans-Planckian censorship
conjecture (TCC)~\cite{TCC} is also satisfied here\footnote{This recent TCC condition
was also recently analyzed in the context of WI in Ref.~\cite{Das:2019hto}.
However, we find that this conjecture (along with the older ones)
requires in fact a much larger dissipation than the $Q\sim 20$ value
estimated in Ref.~\cite{Das:2019hto}. This is also backed up by the model
studied in Ref.~\cite{Bastero-Gil:2019gao}. This is so because the conjecture TCC
requires a very small scale for inflation in general and in WI this can only be
achieved if very large $Q_*$ values are considered, at least for the presently known
and treated models in the literature.}.  Hence, to find that the
results found here can satisfy all previously mentioned constraints
is quite a pleasant feature\footnote{Note also that as recently claimed in Ref.~\cite{Mizuno:2019bxy},
a larger scale for inflation than the one found in~\cite{TCC} might in fact be required 
from the TCC. Though this would lessen the condition on $Q$ here, we still need
a sufficiently large $Q$ so as to have a temperature low enough at the end of inflation to satisfy
the other bounds on the model.}.

There is also another important consistent check that needs to be
verified here.  Since we are working with very steep potentials, and
consequently, also with very efficient and large dissipation, where
$Q_* \gg 1$, one might worry how such a large dissipation might affect
the spectrum beyond the linear order. Since in WI the inflaton and
radiation perturbations get very strongly coupled at large dissipation
ratios, we expect that large nonlinearities will emerge and
adversely affect the spectrum of perturbations. As far the non-Gaussianities in
WI are concerned, the Planck team has produced constraints for the
so-called warm shape of the bispectrum in WI~\cite{Moss:2007cv}, with
the non-Gaussianty coefficient denoted by $f_{\rm NL}^{\rm warm}$. In the
analysis done in Ref.~\cite{Moss:2007cv} and valid for the strong
dissipative regime of WI (but for a temperature independent and constant 
dissipation ratio $Q$), the
expression  found for $f_{\rm NL}^{\rm warm}$, valid in the strong dissipative regime, 
was\footnote{Note that
  in the early Ref.~\cite{Moss:2007cv} and also in the Planck
  papers on the non-Gaussianities, the notation used for the
  dissipation coefficient $Q$ was $r_d$.}  $f_{\rm NL}^{\rm warmS}=-15
\ln (1+Q/14)-5/2$.  The Planck 2018 analysis~\cite{Akrami:2019izv}
based on this expression has given the results,  $f_{\rm NL}^{\rm
  warmS}=-48\pm 27$ (from SMICA+T+E, $68\%$ C.L.) and   $f_{\rm
  NL}^{\rm warmS}=-39\pm 44$ (from SMICA+T, $68\%$ C.L.), which then
can be translated into the upper bounds for the dissipation ratio, ${\rm
  log}_{10} Q \leq 3.5$ and ${\rm log}_{10} Q \leq 3.6$, respectively,
at $95\%$ C.L. The largest values for $Q_*$ that we have and that are
shown in  {}Figs.~\ref{fignsXQlambda} and \ref{fignsXQalpha} and to
still to have a  consistent value for $n_s$, $Q_* \lesssim 10^3$, all
fall safely inside these upper bounds from the non-Gaussianity
analysis.  In particular, using the simple formula for the non-Gaussianity parameter
derived in Ref.~\cite{Moss:2007cv} that was considered by the Planck team,  
we find $|f_{\rm NL}^{\rm warm}|
\simeq 61.5$, when using the parameters of the explicit numerical
example given above, with $Q_* = 700$. Later on, in Ref.~\cite{Moss:2011qc}, 
the authors generalized the computation of $f_{\rm NL}^{\rm warm}$ 
for a temperature dependent $Q$ (and it was also pointed a sign error in the previous work). In
Ref.~\cite{Zhang:2015zta}, the authors made use of the
$\delta N$ formalism to study the non-Gaussianity. However, the $\delta N$ 
formalism better probes the non-Gaussianity of the local shape, while the 
warm shape of WI, valid in the strong dissipative regime, is very weakly 
correlated with the local shape~\cite{Bastero-Gil:2014raa}.  In
Ref.~\cite{Bastero-Gil:2014raa}, a complete analysis of
non-Gaussianity in WI was performed, including the full effect of the
coupling of the inflaton and radiation perturbations, which is
essential to gauge the effect of the large dissipation on the
spectrum. {}Following the numerical procedure explained in
Ref.~\cite{Bastero-Gil:2014raa} and using the parameters values we
have for the cubic form of the dissipation coefficient used in this
work and the numerical example given above, for $Q_* = 700$, we find
the result $|f_{\rm NL}^{\rm warm}| \simeq 5.5$, which is small enough
to satisfy the Planck bounds, but still large enough to possibly be
probed in the future through more precise observations from both fourth
generation CMB observatories and on also future large scale structure surveys,  
which are expected to bring down the present upper bounds on non-Gaussianities.

\section{Conclusions}
\label{concl}

In this work, we have shown that dissipation effects, in the WI
context, assist to achieve a consistent inflationary model with an
exponential potential in the RS II braneworld. It also allows us to
simultaneously satisfy all the theoretical and observational
restrictions given previously on these type of exponential inflation
potential in the braneworld construction. In fact, this is the first
model of warm inflation with the specific form of a cubic temperature
dependent dissipation coefficient that is found to be consistent with
the observations in the strong dissipative regime of WI.

Achieving the strong dissipative regime in the present model is a
consequence of both the braneworld high energy change of the GR
evolution, combined with the exponential form for the primordial
inflaton potential.  As the result of achieving such a consistent
dynamics in the strong dissipative regime, where $Q \gg 1$, it turns
out that one can easily satisfy all the swampland conditions that have
been recently proposed. Moreover, the model has a graceful exit from
inflation for a large value of the constant parameter $\alpha$ in the
inflaton exponential potential, with the dissipation ratio $Q$
decreasing and the brane correction disappearing at the end of inflation.
{}Furthermore, although the brane correction can make the
tensor-to-scalar ratio $r$ larger in the high energy level, due to the
change in the tensor spectrum, the dissipation effects suppress the
energy scale of inflation more significantly in the strong dissipative
regime and the model turns out to be fully consistent with
observational data, with appropriate values for the spectral tilt and
a highly suppressed tensor-to-scalar ratio, which is one of the main
results of the present work. Additionally, the temperature of the
Universe at the end of inflation is inside the range that the model
does not suffer from potential gravitino overproduction and we can
also avoid a large amount of relic gravity waves. 
We also note that these results follow for a broad range of the model parameter
values, like for the constant $\alpha$ in the inflaton potential and
for the brane tension $\lambda$, yet always satisfying the constrain bounds 
for these parameters. As also seen from the results shown in {}Figs.~\ref{fignsXQlambda}
and \ref{fignsXQalpha}, there is a broad range of values for the dissipation 
coefficient that can be found to be consistent with the Planck data.
In this sense, there is no special imposition or need for specific fine-tunings on
these parameters. Likewise, as we study a regime of strong dissipation in WI,
the inflationary slow-roll trajectory can be approached much faster, which
reinforces the attractor like behavior of the corresponding slow-roll 
solution (see, e.g., Refs.~\cite{deOliveira:1997jt,Ramos:2001zw} for earlier
studies of the effect of dissipation on the inflationary slow-roll trajectories).

We recall that exponential type of potentials have been considered as
a possible candidate for describing the late-time
acceleration~\cite{Copeland:1997et,LopesFranca:2002ek}
observed in the recent Universe.  However, by having a tracking
behavior,  it fails to properly act in the present-time as dark
energy, with an equation of state that must be $\omega_\phi \simeq -1$.
Besides, even in the context of the braneworld scenario, problems with
this type of potential has been exposed~\cite{Dias:2010rg}, like a too
large tensor-to-scalar ratio, a too red-tilted spectral index, and
a possible excess of gravitational waves production that can destroy the
BBN. {}From the results we have presented in this
work, we see that all these problems  can be overcome. {}Furthermore,
typical interactions in the dark sector recently proposed in
Ref.~\cite{Lima:2019yyv}, and fully motivated from WI, could also be used in
the present context.  Since the Universe can smoothly enter into the
radiation dominated epoch due to the presence of the WI dissipative
effects, the reheating regime can be evaded, and the remaining inflaton
field can be thawed and can behave like quintessence at latetime,
like in the models studied in Ref.~\cite{Lima:2019yyv}. Hence, the
exponential potential can still be practical for unifying both the recent
and the very early Universe histories, in the so-called
dissipative quintessential inflation scenario.  Therefore, our model can be a
first step towards reconciling inflation and dark energy, with the
bonus of being consistent with a high-energy UV completion within a
theory of quantum gravity. We will study this possibility in the
near future as a separate work.

\section*{Acknowledgments}

V.K's research at McGill has been supported by a NSERC Discovery Grant
to Robert Brandenberger.  R.O.R. is partially supported by research
grants from Conselho Nacional de Desenvolvimento Cient\'{\i}fico e
Tecnol\'ogico (CNPq), Grant No. 302545/2017-4, and Funda\c{c}\~ao
Carlos Chagas Filho de Amparo \`a Pesquisa do Estado do Rio de Janeiro
(FAPERJ), Grant No. E-26/202.892/2017.


\end{document}